\documentclass[10pt]{iopart}
\pdfoutput=1

\usepackage{cite}
\usepackage{color}
\usepackage{graphicx}
\usepackage{notes2bib}
\usepackage{float}

\usepackage{ae}
\usepackage{bm}

\usepackage{tikz}
\definecolor{pastell}{rgb}{0.066,0.289,0.519}
\definecolor{darkred}{rgb}{0.585,0,0}
\definecolor{darkgreen}{rgb}{0,0.585,0}
\definecolor{darblue}{rgb}{0,0,0.585}
\definecolor{lightmagenta}{rgb}{1,0,1}
\definecolor{velvet}{rgb}{0.468,0.13,0.4}
\definecolor{redblue}{rgb}{0.5,0,0.5}
\definecolor{redgreen}{rgb}{0.5,0.5,0}
\definecolor{greenblue}{rgb}{0,0.5,0.5}
\definecolor{maroon}{cmyk}{0,0.87,0.68,0.32}

\newcommand{\text}{\mathrm}

\newcommand{\Tc}{\ensuremath{T_\mathrm{c}}}

\newcommand{\Tg}{T_{\text{g}}}

\newcommand{\eqref}[1]{~(\ref{#1})}

\newcommand{\vs}{v_{\text {s}}}
\newcommand{\lc}{l_{\text {c}}}

\newcommand{\hide}[1]{}

\usepackage[normalem]{ulem}

\renewcommand{\vec}[1]{\bi{#1}}

\begin{document}

\title[A crossover in non-affine response]{A crossover in spatio-temporal correlations of strain fluctuations in glass forming liquids}

\author{Muhammad Hassani$^1$, Marian Bruns$^1$, Fathollah Varnik$^1$}

\address{$^1$Interdisciplinary Centre for Advanced Materials Simulation (ICAMS), Ruhr-Universit\"at Bochum, Universit\"atsstra{\ss}e 150, 44801 Bochum, Germany}
\ead{fathollah.varnik@rub.de}
\vspace{10pt}
\begin{indented}
\item[]1 March 2016
\end{indented}

\begin{abstract}
Via molecular dynamics simulations of a generic glass former in the supercooled and normal liquid states, it is shown that spatial correlations of strain fluctuations exhibit a crossover from the well-established power-law $\sim 1/r^3$-decay at long wavelengths to an exponential behavior, $\sim \exp(-r/l_{\text {c}})$ at intermediate distances. The characteristic length of the exponential decay grows both with temperature and time via, $l_{\text {c}}^2 \propto D(T)\, t$, with $D(T)$ being the temperature-dependent diffusion coefficient. This suggests that the crossover between the power-law and exponential decays is governed by a diffusion process.
\end{abstract}


\section{Introduction}
\label{sec:introduction}
Based on their properties such as high corrosion resistance and strength, glass formers and in particular glassy alloys have attracted substantial attention among physicist and material scientists~\bibnote{As an example, in Multiscale Material Modeling (MMM) conference 2018 in Osaka, which is a diverse multi-disciplinary conference, two (out of six) of the plenary talks were on the mechanics of deformation in amorphous materials.}. One of the intriguing features of glassy systems is their non-affine response, which is at the same time a challenging issue to explore mostly 
due to their structure, incorporating no long-range translational order. Glass transition is characteristically very dissimilar to crystallization since it involves no significant change in the fluid-like structure in the quench process, however it slows down considerably the dynamics of particle rearrangements. With the slowed dynamics, the particles can carry forces in their cages, resulting in an elastic response, during which the lack of symmetry in the structure of the glass results in non-affine motions of particles to conserve  momentum~\cite{Goldenberg2007,Falk1998,Royall2015}.

Due to the lack of crystalline order, concepts such as dislocations and glide planes are not applicable in the case of amorphous materials. In this context, the study of non-affine response provides an interesting alternative. In this regard, spatio-temporal correlations of local deviations from the affine response, quantified by $D^{2}_\text{min}$ parameter~\cite{Falk1998}, under continuous loading have provided insight on the accommodation of deformation around plasticity carriers (shear transformation zones or STZs), which control the macroscopic failure of glasses via shear banding~\cite{Nicolas2014,Chikkadi2011, Mandal2013b,Varnik2014, Hassani2016,Hassani2018}. The study of the correlations has been further confirmed by the quasi-static athermal simulations, where the dynamics are slow enough to capture the deformation associated with each stress drop~\cite{Tanguy2006,Maloney2006a,Lemaitre2009,Dasgupta2013,Lagogianni2018}. Both approaches reveal a deformation pattern with quadrupolar symmetry and a power-law decay in long distances.

The qualitative resemblance of the deformation around each local shear events and the Eshelby inclusion problem has been suggested by numerous experiments and simulations~\cite{Eshelby1957,Picard2004,Nicolas2015,Schall2007}. This resemblance highlights the role of elasticity in mediating the deformation around each event.  In contrary to this presumption about the role of elasticity, there have been reports on the long-range strain correlations in supercooled liquids, where no high-frequency elasticity could be defined, under shear~\cite{Chattoraj2013} and also in quiescent condition~\cite{Illing2016}. Recent theoretical development via mode coupling theory (MCT) and generalized hydrodynamics (GH) have proposed that a wave-length-and-frequency-dependent response in liquids results in a solid-like region around a local shear perturbation, expanding up to a length which a wave traverse with the speed of sound during the corresponding structural relaxation time~\cite{Illing2016,Hassani2018b}. Moreover, similar spatial characteristics have been also reported in correlations of stress fluctuations~\cite{Lemaitre2014,Lemaitre2015}.

While the developed theories, simulations and experiments mostly have addressed strain~\cite{Illing2016,Hassani2018b,Chattoraj2013} and stress~\cite{Maier2017,Lemaitre2017,Lemaitre2018} correlations in the asymptotic limit of large distances, in the present work, a distinction is made between the response at long distances and the one in the intermediate range. Most importantly, we show that, in the liquid state, the well-established power-law decay of the correlations of non-affine strain at long distances crosses over to an exponential behavior at intermediate lengths. For all the temperatures investigated, the characteristic length associated with this new type of decay approaches a diffusive growth with time, $\lc\propto D(T)\, t$, with $D(T)$ being the temperature-dependent diffusion coefficient.

\section{Method}
The Kob-Andersen model~\cite{Kob1994}, as a generic model for glass formers, is used in our three dimensional molecular dynamics (MD) simulations. Via this model, various issues have been investigated, such as non-Newtonian rheology \cite{Berthier2002a,Varnik2006d}, heterogeneous plastic deformation and flow \cite{Varnik2003,Hassani2016} and structural relaxation under shear \cite{Berthier2002a,Varnik2006b}. The model contains two types of atoms, $\text{A}$ and $\text{B}$ ($80:20$), which interact with the Lennard-Jones (LJ) potential (6-12) and have diameters of $d_\text{AA}=1$, $d_\text{BB}=0.8$, and $d_\text{AB}=0.88$. The interaction energy between the two types of particles are $\epsilon_\text{AA} = 1.$, $\epsilon_\text{BB}=0.5$, $\epsilon_\text{AB}=1.5$. The total mass density of $\rho=\rho_\text{A}+\rho_\text{B}=1.2$ results in $\sim1.2\times 10^6$ particles ($m_\text{A}=m_\text{B}=1$) in a cubic box with sides of $L=100$. The simulation box is periodic in all three spatial directions. All the quantities are normalized with the attributes of particle $\text{A}$ and "$\text{A-A}$" LJ interaction, such as $d_\text{AA}$, $m_\text{A}$, and $\epsilon_\text{AA}$. The simulations are performed using LAMMPS~\cite{Plimpton1995} with a time step of $\delta t = 0.005$. 

The investigation of the correlations is performed via quiescent simulations, where the non-affine rearrangements of particles are due to thermal fluctuations~\cite{Rodney2009}. Glassy state of $T=0.2$, supercooled state of $T=0.7$, and liquid states of $T=1.5$, $2.0$, and $4.0$ are studied.
The mode coupling critical temperature of the model is $\Tc=0.435$~\cite{Kob1995} and the glass transition temperature, $\Tg \sim 0.4$~\cite{Varnik2006d}. The high-temperature liquid states are chosen for this study so that with the help of higher diffusion, the dynamics of the proposed crossover from the intermediate to long-distance correlations are better highlighted.   

\subsection{Correlations of strain fluctuation}
To calculate the local strain, at a position $\vec{r}_0$ and a time $t_0$, accumulated over a time interval of $t$, $\epsilon(\vec{r}_0,t_0,t)$, the trajectories of the particles in a neighborhood are analyzed. The displacement of each particle (e.g., $i$), $\vec{u}_i(t)=\vec{r}_i(t_0+t)-\vec{r}_i(t_0)$ is coarse grained (CG), on its neighborhood via a distance-dependent exponential weighting function of $\phi(||\vec{r}_0-\vec{r}_{i}||)$, to $\vec{u}^\text{CG}(\vec{r}_0)$ \cite{Goldenberg2007}.

In our quiescent simulations, the displacement, $\vec{u}_i$ is per se non-affine and is due to thermal fluctuations. The resulting local strain is obtained via $\epsilon(\vec{r}_0,t_0,t)=\frac{1}{2}\left[\nabla \vec{u}^{CG} +\left(\nabla \vec{u}^{CG}\right)^{\top} \right]$.

\begin{figure}
	\begin{center}
		\includegraphics[height=7cm]{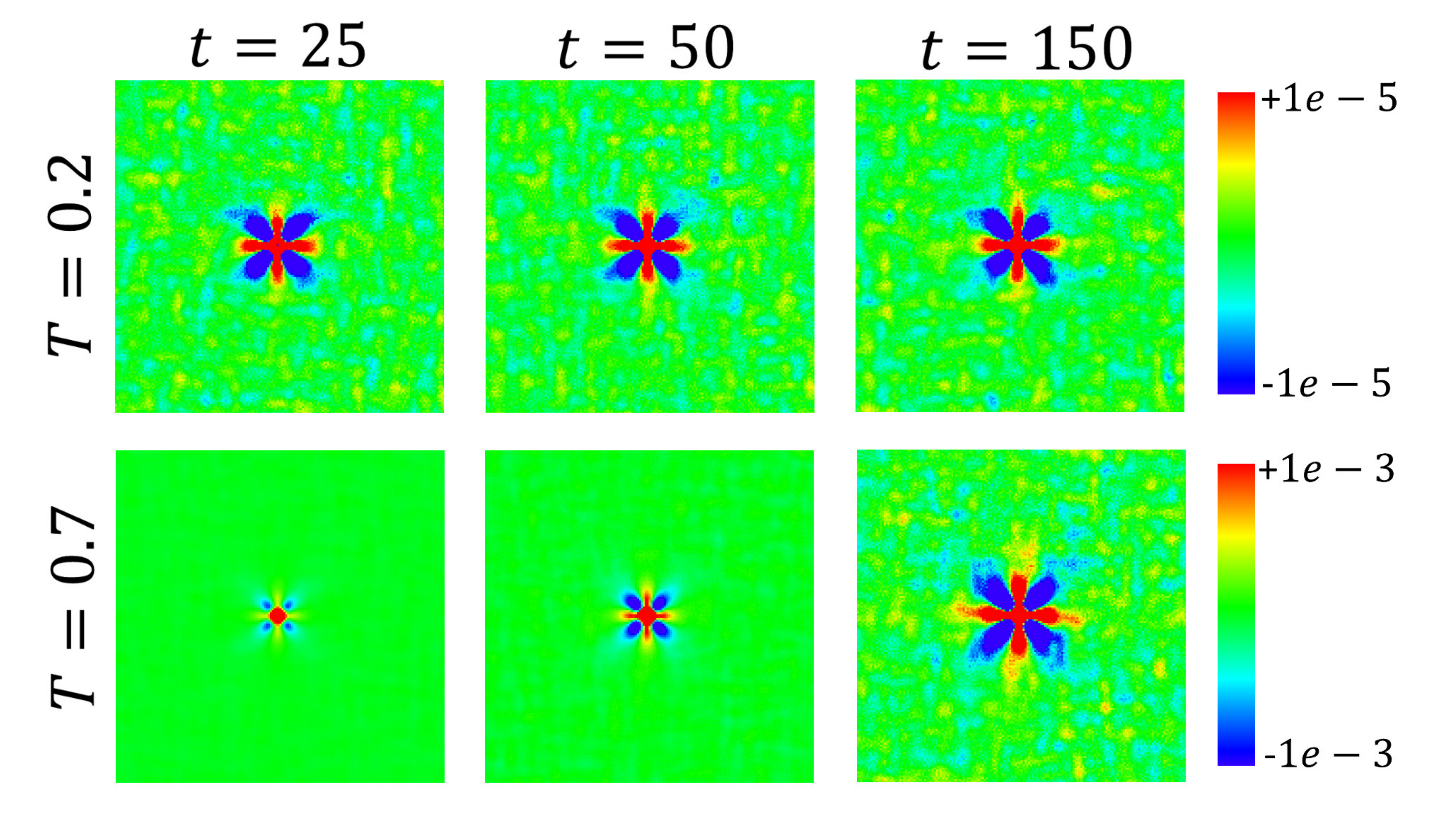}
	\end{center}
	\caption{Temporal evolution of the correlations $ C_{\mathbf{\epsilon}}(\vec{r},t)$ in the bulk at the glassy state ($T=0.2$, top row), and supercooled state ($T=0.7$, lower row). In both states and in all the time intervals shown, the quadrupolar symmetry is preserved. At the glassy state, as the time intervals increases, correlations hardly change, while in the supercooled state, the correlations grow in time (Please note the different scales for the color map in the two cases).}
	\label{fig:2DCorrBulk}
\end{figure}

The spatial correlations of the accumulated strain, between the points $\vec{r}_0$ and $\vec{r}+\vec{r}_0$ are obtained by $C_{\epsilon_{xz}}(\vec{r},t)=\left\langle \epsilon_{xz}(\vec{r}_0+\vec{r},t_0,t)\epsilon_{xz}(\vec{r}_0,t_0,t)\right\rangle$. In our three dimensional MD box, the correlations are calculated within thin layers of one particle diameter thickness. If the thin layer used for this calculation is in the corresponding shearing plane (e.g., in the case of $\epsilon_{xz}$, $xz$-planes), one can expect an angular anisotropy like the patterns in Fig.~\ref{fig:2DCorrBulk}. Therefore, in order to focus only on the distance and time dependence of the correlations, one can use the spherical harmonic projection, defined as $C^4_4(r,t)=\frac{1}{\pi}\int_{0}^{2\pi}{C_{\epsilon_{xz}}(\vec{r},t)\cos(4\theta)d\theta}$. Here, $r$ and $\theta$ are chosen as $\vec{r}=r\left(\cos(\theta),0,\sin(\theta)\right)$. In the quiescent simulations, by taking the advantage of the isotropy of the system in the absence of external deformations, all the three non-diagonal components of the strain tensor (i.e. $\epsilon_{xy}$, $\epsilon_{xz}$, and $\epsilon_{yz}$) are exploited to evaluate $C^4_4(r,t)$ in their corresponding shear plane.

\section{Results and discussion}

In this section, we first review the behavior of correlations at long distances and then switch to the novel aspect which deals with the crossover behavior
at intermediates lengths.

\subsection{Power-law behavior at large distances}
For large distances (small wave vectors), the generalized hydrodynamic (GH) theory provides predictions for the correlation of strain fluctuations (Eqs.~(5) and~(6) in~\cite{Hassani2018b}) as,
\begin{equation}
C_{\epsilon_{xz}}(r,t)=\frac{3}{8\pi} \left(\frac{v_{th}}{\vs}\right)^2\frac{J_M(t)}{r^3} \frac{r^2(x^2+z^2)-10x^2z^2}{r^4};\;\; a_{\text{micro.}}<r<\xi,
\label{eq:corr}
\end{equation} 
where $v_\text{th}$ and $\vs$ are thermal velocity and sound speed, respectively, and $J_M$ is creep compliance. The correlations are predicted within an elastic domain of size $\xi$, and for distances large compared to a microscopic length size, $a_{\text{micro.}}$; the extent of the latter will be discussed in the next section.

Equation~(\ref{eq:corr}) reveals a power-law decay of $1/r^3$, where the term $10x^2z^2$ expresses the four-fold symmetry in the correlations. This strongly resembles the strain field around a pre-sheared spherical inclusion in continuum elasticity (Eshelby problem)~\cite{Eshelby1957}.

By applying the spherical harmonic projection operator to Eq.\eqref{eq:corr} and assuming the vector $\vec{r}$ to be parallel to $xz$-plane, one obtains

\begin{equation}
C^4_4(r,t)=\frac{1}{\pi}\int_0^{2\pi}{C_{\epsilon_{xz}}(r,t)\cos(4\theta) d\theta}=\frac{C_s(t)}{r^3},
\label{eq:C4}
\end{equation}
where the amplitude $C_s(t)$, by asymptotic analysis of $J_M(t)$, reads~\cite{Hassani2018b},
\begin{equation}
C_s(t)= \cases{
	\frac{15\rho k_\text{B} T}{32 \pi m}\Big(\frac{1}{G^\perp_{\infty}}-\frac{1}{G^\Vert_{\infty}}\Big), \;\;\; t\ll\tau \\
	\frac{15\rho k_\text{B} T}{32 \pi m}\frac{t}{\eta},\;\;\;\;\; t\gg\tau.
}
\label{eq:Cs}
\end{equation}

\begin{figure}
	\hspace*{-1.cm}
	\includegraphics[height=6.5cm]{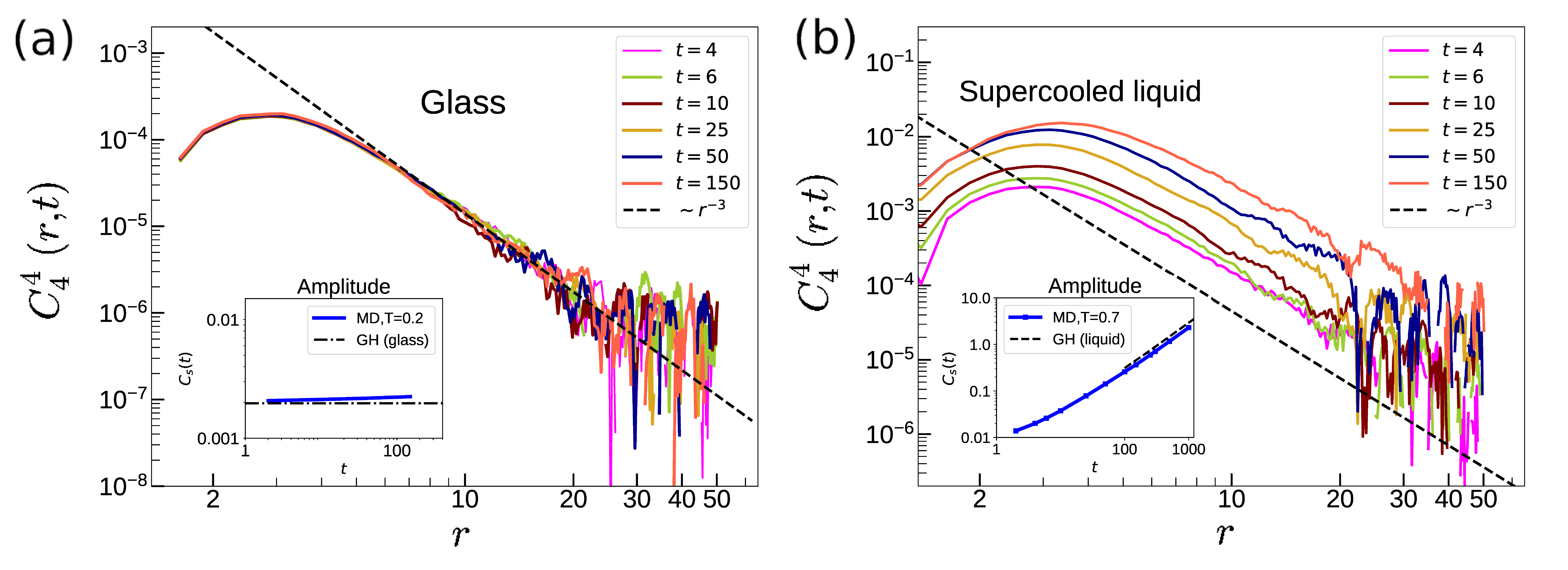}
	\caption{Spatio-temporal correlation of strain fluctuations, $C_4^4(r,t)$, obtained from MD simulations of a generic binary LJ glass former at temperatures of (a) $T=0.2$ (glassy state) and (b) $T=0.7$ (supercooled liquid). For both states, the correlations depict a power-law decay of $1/r^3$, at large distances. Growing correlations in the supercooled state and non-evolving correlations in the glassy state in the two panels are consistent with the patterns in Fig.~\ref{fig:2DCorrBulk}. In the insets of both panels, the amplitude of the correlations, $C_s(t)$ is displayed and compared with the predictions from generalized hydrodynamic theory (GH), Eq.\eqref{eq:Cs}. In quantitative agreement with GH, the amplitude in the glassy state depicts a quasi-plateau with time, whereas a linear growth of the amplitude is observed in the supercooled state. The parameters used in the evaluation of theoretical predictions,Eq.\eqref{eq:Cs}, are $G^\perp_\infty(T=0.2)=15$, $G^\Vert_\infty(T=0.2) = 86$, $\eta(T=0.7)=42$, and $\rho/m =1.2$.}
	\label{fig:C4-0.2N0.7}
\end{figure}

In Eq.\eqref{eq:Cs}, $\eta$ represents the shear viscosity of the liquid; $G^\perp_{\infty}$ and $G^{||}_\infty$ correspond to the transverse and longitudinal shear moduli of the glass at intermediate frequencies (the frequencies associated with the time intervals of the plateau in the mean squared displacement).

From the quiescent three dimensional MD simulations, the spatio-temporal correlations of strain fluctuations are evaluated. Figure~\ref{fig:2DCorrBulk} illustrates a color-scale plot of the thus obtained results both for the glassy ($T=0.2$) and supercooled ($T=0.7$) states. In both cases shown, correlations display quadrupolar symmetry. However, while the four-fold correlations in the glassy state do not change noticeably with time, growth is visible in the supercooled state.

For an analysis of the distance dependence and of the temporal evolution of these correlations, we project the two-dimensional data onto a spherical harmonics, Eq.\eqref{eq:C4}. This averaging over the angles allows a more telling quantitative survey of the data. The thus obtained curves are shown in Fig.~\ref{fig:C4-0.2N0.7}.
\begin{figure}
	\hspace*{-1.5cm}
	\includegraphics[height=6.5cm]{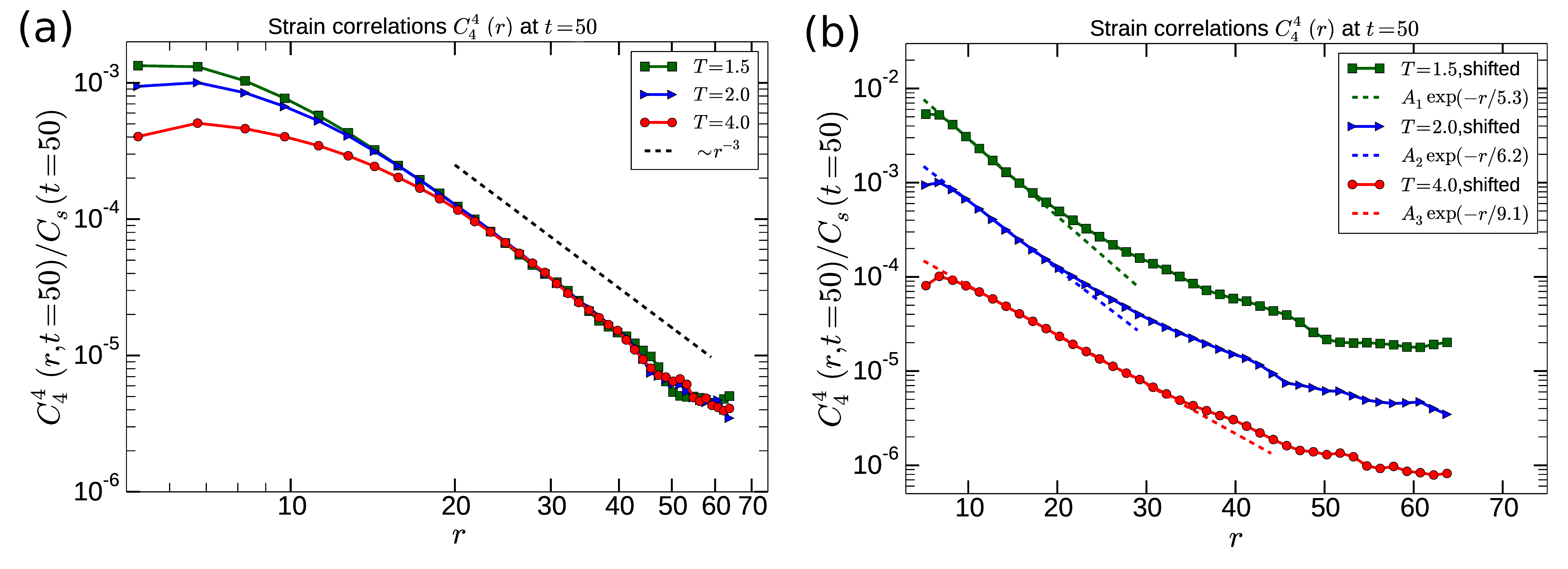}
	\caption{(a) Logarithmic plot of normalized strain correlations, $C_4^4(r,t)/C_s(t)$ at various temperatures: $T=1.5,\text{ }2.0,\text{ }4.0$ and a single time interval of $t=50$. Despite the high temperatures, in large distances, the correlations decay as a power-law function of $r^{-3}$. (b) Semi-logarithmic plot of the same data as in (a) (vertically shifted for a better visibility). This plot serves to highlight the fact that, for all three temperatures investigated, the correlations exhibit an exponential decay, $\sim \exp(-r/\lc)$ at intermediate distances. It is also visible from this plot that the characteristic length, $\lc$, grows with temperature.}
	\label{fig:diffTemp-exp}
\end{figure}

At the glassy state (Fig.~\ref{fig:C4-0.2N0.7}-a), for all the time intervals, correlations follow a master curve with a power-law decay of $1/r^{3}$. Note that the data are not rescaled here, indicating that correlations do not substantially change with time in the glass. This is consistent with the stationary quadrupolar patterns in Fig.~\ref{fig:2DCorrBulk}. The amplitude of these correlations, $C_s(t)$, is obtained via fits of the form $C_s/r^{3}$ to the simulation results on $C_4^4(r,t)$ for each individual value of time, $t$. In this fit, we focus on the asymptotic behavior and neglect the part of the data at too short distances, with a cut-off roughly of the order of the decay range of pair correlation function. As shown in the inset of Fig.~\ref{fig:C4-0.2N0.7}-a, the thus obtained $C_s(t)$ agrees well with the prediction of the generalized hydrodynamic theory, Eq.\eqref{eq:Cs}. It is noteworthy that $G^\perp_{\infty}$ and $G^\Vert_{\infty}$ used in this comparison are not treated as free fit parameters but are found to be $G^\perp_\infty(T=0.2)=15$ and $G^\Vert_\infty(T=0.2) = 86$ from independent simulations of oscillatory shear and then used without any modification~\cite{Hassani2018b}. This underlines the physical consistency of the test.

Similar to the glassy state, $C_4^4(r,t)$ evaluated in the supercooled state shows a $1/r^3$ power-law decay for large distances, Fig.~\ref{fig:C4-0.2N0.7}-b. There is, however, an important difference. While the amplitude of correlations remains essentially constant with time in the glass, it grows linearly with time in the supercooled liquid state. This linear growth is quantified in the inset of Fig.~\ref{fig:C4-0.2N0.7}-b and agrees well with the prediction of generalized hydrodynamics. Again, the viscosity used in this comparison is not a fit parameter but obtained from independent steady shear simulations ($\eta(T=0.7)=42$).

\begin{figure}
	\begin{center}
		\includegraphics[width=13cm]{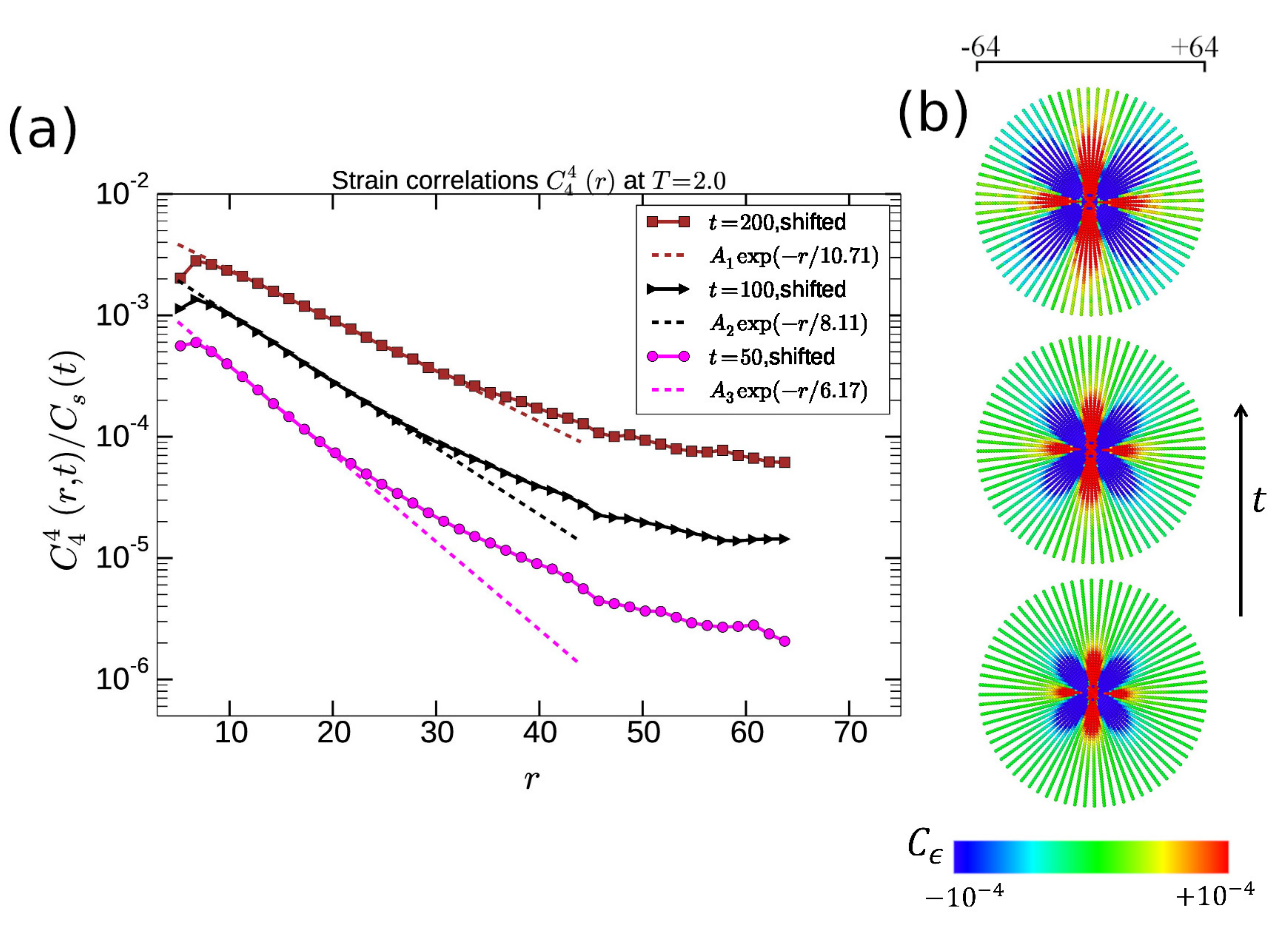}
	\end{center}
	\caption{Left panel: spatio-temporal correlations of strain fluctuations, $C_4^4(r,t)$, plotted in a semi-logarithmic scale at temperature of $T=2.0$, evaluated for three time intervals of $t=50,\text{ }100, \text{and}~200$. The correlations in all three time intervals depict an initial exponential decay. The size of the exponential region expands by time which could reach up to 40 particle diameters for the case of $t=200$. Right panel: 2D spatial evolution of the correlations are shown. Even though the exponential decay extends up to 40 particle diameters, the four-fold pattern of the correlations is preserved.}
	\label{fig:temporalExp}
\end{figure}

\begin{figure}
	\includegraphics[height=6.0cm]{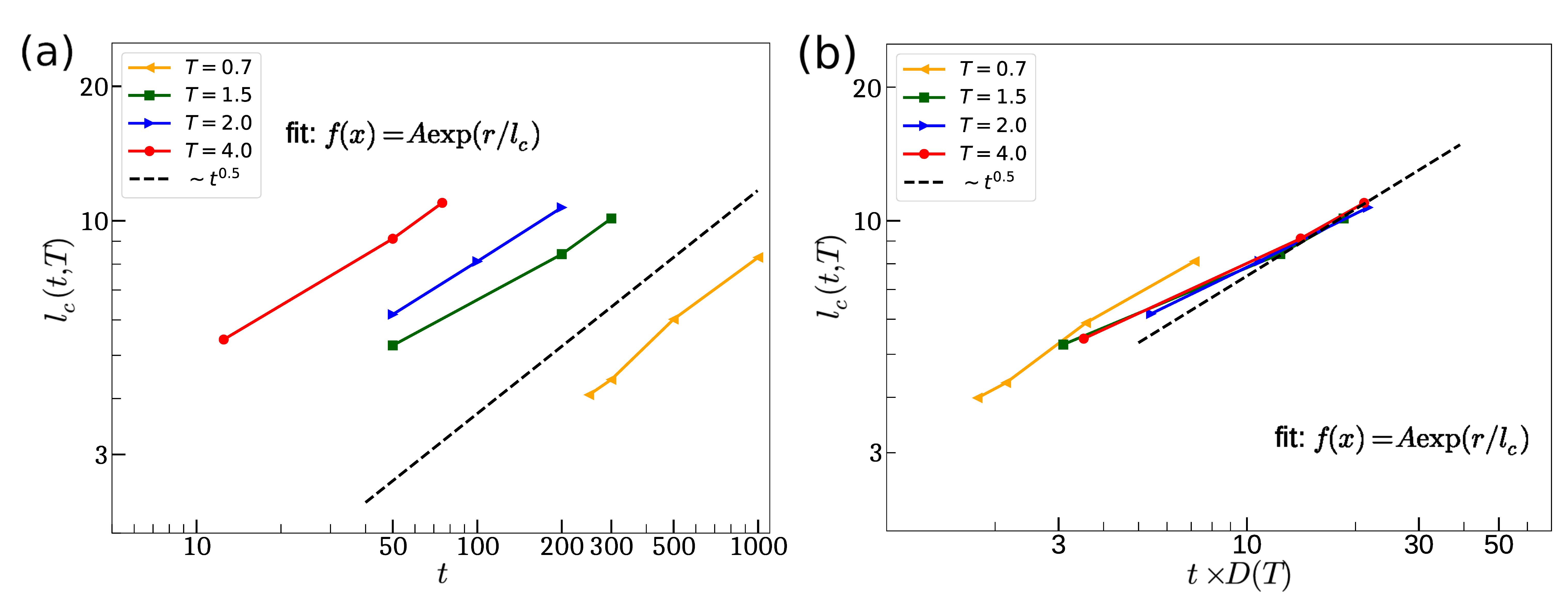}
	\caption{(a) Time dependence of the characteristic length, $\lc$, of the exponential decay of the non-affine strain correlations, $C_4^4(r,t)$, for a number of temperatures as indicated. The lowest temperature ($T=0.7$) belongs to the supercooled state, while all other temperatures shown here belong to the normal liquid state, where no signature of a two-step relaxation exists. The dashed line is a guide to the eye and serves to highlight a quasi-diffusive growth of $\lc$. (b) The same data as in (a), now plotted versus the product of time and diffusion coefficient. In this representation, the data corresponding to different temperatures follow approximately a  master curve. Again, the dashed line indicates diffusive behavior. Diffusion coefficients used in this plot are extracted from the slope of mean square displacements versus time in the limit of long times. The values of $D$ used here are obtained independently from mean square displacements and read $D(T=0.7)=0.0072$, $D(T=1.5)=0.062$, $D(T=2.0)=0.11$, and $D(T=4.0)=0.28$.}
	\label{fig:lengthScale}
\end{figure}

\subsection{Exponential decay at intermediate lengths}
It is shown above that, for long distances, the correlations of non-affine strain obtained from our simulations, support quantitatively the existence of a long range power-law decay, characteristic of an elastic body. Here, we show that the same correlation functions exhibit a qualitatively different behavior for intermediate distances and that the size of this "intermediate range" is temporally-evolving and temperature-dependent. There thus is a crossover between two different spatial and temporal dependencies of the correlations of non-affine strain.

As a first piece of evidence for the existence of a crossover, Fig.~\ref{fig:diffTemp-exp} shows the normalized correlation function, $C_4(r,t)/C_s(t)$, versus distance for a time interval of $t=50$ but different temperatures, all belonging to the liquid state. In order to highlight the different decay-laws at intermediate and long distances, the same data are plotted in two different ways. While a logarithmic plot help to identify a possible power-law behavior easily, a semilogarithmic representation makes it easier to judge about a possible exponential decay. As seen from the corresponding panels of Fig.~\ref{fig:diffTemp-exp}, the correlation of non-affine strain first follows an exponential decay and then switches over to the well-known power-law at longer distances~\cite{Hassani2018,Hassani2018b}. 

It is interesting that this behavior is clearly present at relatively high temperatures belonging to the normal liquid state. It is also visible from this figure that the spatial range over which the data follow an exponential dependence increases with temperature. We will see later below that this is closely related to an increase of diffusion coefficient with $T$.

To explore this issue further, Fig.~\ref{fig:temporalExp}-a focuses on the temporal evolution of the exponential decay at a temperature of $T=2.0$ (liquid state). For this purpose, the data are shown for time intervals of $t=50$, $100$, and $200$. It is visible from the shown data that the range over which an exponential fit describes the data grows with time. It is also noteworthy that the four-fold pattern survives the crossover from power-law to exponential behavior as it is clearly present in Fig.~\ref{fig:temporalExp}-b.

Combining the observations from Figs.~\ref{fig:diffTemp-exp} and Fig.~\ref{fig:temporalExp}, one sees that the characteristic length for the exponential decay,  $\lc$, grows both with temperature and time. A closer survey of this length is provided in Fig.~\ref{fig:lengthScale}a, where $\lc$ is plotted versus time and seems to follow a diffusive time dependence, in a way reminiscent of mean square displacements at long times. To check this idea, Fig.~\ref{fig:lengthScale}b represents the same data as function of $D\,t$, the product between diffusion coefficient and time, revealing an approximate master curve.

\section{Conclusion and outlook}

For a wide range of temperatures, ranging from the normal liquid to the supercooled state of a simple model glass, correlations of strain fluctuations show a cross-over from a long-range power-law dependence, $~1/r^3$, typical of a homogeneous and isotropic elastic body, to an exponential decay at intermediate distances. The characteristic length associated with this new behavior is not constant but grows with time as $\lc\propto D(T)\,t$, where diffusion coefficient $D(T)$ entails the temperature dependence of $\lc$. This is strongly reminiscent of the growth of mean square displacements with time in the diffusive regime. To date, we are not aware of any theory describing this observation.

The present observation of a crossover in the correlations of strain fluctuations from a power-law decay to an exponential behavior both in  the normal liquid and in the supercooled state is of fundamental importance for a better understanding of plastic deformation in amorphous materials and calls for new theoretical work. A possible route here could be to extend the recently proposed generalized hydrodynamics approach~\cite{Hassani2018b} to finite wave vectors.

\section{Acknowledgments}
Financial support by the German Research Foundation (DFG) under the grant number VA 205/18-1 is acknowledged.

\section*{References}
	
\end{document}